\documentclass[useAMS,usenatbib]{mn2e}

\usepackage{xspace}
\usepackage{graphicx}
\usepackage{tabularx}
\newcommand{\ignore}[1]{}
\providecommand{\ao}{}
\renewcommand{\ao}{adaptive optics (AO)\renewcommand{\ao}{AO\xspace}\renewcommand{\Ao}{AO\xspace}\xspace}
\newcommand{\Ao}{Adaptive optics (AO)\renewcommand{\ao}{AO\xspace}\renewcommand{\Ao}{AO\xspace}\xspace}
\newcommand{\wfs}{wavefront sensor (WFS)\renewcommand{\wfs}{WFS\xspace}\renewcommand{\wfss}{WFSs\xspace}\xspace}
\newcommand{\wfss}{wavefront sensors (WFSs)\renewcommand{\wfs}{WFS\xspace}\renewcommand{\wfss}{WFSs\xspace}\xspace}
\newcommand{\shwfs}{Shack-Hartmann \wfs (SHWFS)\renewcommand{\shwfs}{SHWFS\xspace}\xspace}
\newcommand{\dm}{deformable mirror (DM)\renewcommand{\dm}{DM\xspace}\renewcommand{\dms}{DMs\xspace}\renewcommand{\Dms}{DMs\xspace}\renewcommand{\Dm}{DM\xspace}\xspace}
\newcommand{\dms}{deformable mirrors (DMs)\renewcommand{\dm}{DM\xspace}\renewcommand{\dms}{DMs\xspace}\renewcommand{\Dms}{DMs\xspace}\renewcommand{\Dm}{DM\xspace}\xspace}
\newcommand{\Dms}{Deformable mirrors (DMs)\renewcommand{\dm}{DM\xspace}\renewcommand{\dms}{DMs\xspace}\renewcommand{\Dms}{DMs\xspace}\renewcommand{\Dm}{DM\xspace}\xspace}
\newcommand{\Dm}{Deformable mirror (DM)\renewcommand{\dm}{DM\xspace}\renewcommand{\dms}{DMs\xspace}\renewcommand{\Dms}{DMs\xspace}\renewcommand{\Dm}{DM\xspace}\xspace}

\newcommand{\shs}{Shack-Hartmann sensor (SHS)\renewcommand{\shs}{SHS\xspace}\renewcommand{\shss}{SHSs\xspace}\xspace}
\newcommand{\shss}{Shack-Hartmann sensors (SHSs)\renewcommand{\shs}{SHS\xspace}\renewcommand{\shss}{SHSs\xspace}\xspace}
\newcommand{\lgs}{laser guide star (LGS)\renewcommand{\lgs}{LGS\xspace}\renewcommand{\lgss}{LGSs\xspace}\xspace}
\newcommand{\lgss}{laser guide stars (LGSs)\renewcommand{\lgs}{LGS\xspace}\renewcommand{\lgss}{LGSs\xspace}\xspace}
\newcommand{\ngs}{natural guide star (NGS)\renewcommand{\ngs}{NGS\xspace}\renewcommand{\ngss}{NGSs\xspace}\xspace}
\newcommand{\ngss}{natural guide stars (NGSs)\renewcommand{\ngs}{NGS\xspace}\renewcommand{\ngss}{NGSs\xspace}\xspace}
\newcommand{\mems}{Micro-Electro-Mechanical Systems (MEMS)\renewcommand{\mems}{MEMS\xspace}\xspace}
\newcommand{\snr}{signal to noise ratio (SNR)\renewcommand{\snr}{SNR\xspace}\xspace}
\newcommand{\moao}{multi-object \ao (MOAO)\renewcommand{\moao}{MOAO\xspace}\xspace}
\newcommand{\mcao}{multi-conjugate adaptive optics (MCAO)\renewcommand{\mcao}{MCAO\xspace}\xspace}
\newcommand{\ltao}{laser tomographic adaptive optics (LTAO)\renewcommand{\ltao}{LTAO\xspace}\xspace}
\newcommand{\cpu}{central processing unit (CPU)\renewcommand{\cpu}{CPU\xspace}\renewcommand{\cpus}{CPUs\xspace}\xspace}
\newcommand{\cpus}{central processing units (CPUs)\renewcommand{\cpu}{CPU\xspace}\renewcommand{\cpus}{CPUs\xspace}\xspace}
\newcommand{\psf}{point spread function (PSF)\renewcommand{\psf}{PSF\xspace}\renewcommand{\psfs}{PSFs\xspace}\xspace}
\newcommand{\psfs}{point spread functions (PSFs)\renewcommand{\psf}{PSF\xspace}\renewcommand{\psfs}{PSFs\xspace}\xspace}
\newcommand{\fpga}{field programmable gate array (FPGA)\renewcommand{\fpga}{FPGA\xspace}\renewcommand{\fpgas}{FPGAs\xspace}\xspace}
\newcommand{\fpgas}{field programmable gate arrays (FPGAs)\renewcommand{\fpga}{FPGA\xspace}\renewcommand{\fpgas}{FPGAs\xspace}\xspace}
\newcommand{\sor}{successive over-relaxation (SOR)\renewcommand{\sor}{SOR\xspace}\xspace}
\newcommand{\fdpcg}{Fourier domain pre-conditioned gradient (FDPCG)\renewcommand{\fdpcg}{FDPCG\xspace}\xspace}
\newcommand{\map}{maximum a-posteriori (MAP)\renewcommand{\map}{MAP\xspace}\xspace}
\newcommand{\elt}{Extremely Large Telescope (ELT)\renewcommand{\elt}{ELT\xspace}\renewcommand{\elts}{ELTs\xspace}\xspace}
\newcommand{\elts}{Extremely Large Telescopes (ELTs)\renewcommand{\elt}{ELT\xspace}\renewcommand{\elts}{ELTs\xspace}\xspace}
\newcommand{\dugall}{Durham University generalised adaptive optics laser laboratory (DUGALL)\renewcommand{\dugall}{DUGALL\xspace}\xspace}
\newcommand{\fwhm}{full-width at half-maximum (FWHM)\renewcommand{\fwhm}{FWHM\xspace}\xspace}
\newcommand{\wht}{William Herschel Telescope (WHT)\renewcommand{\wht}{WHT\xspace}\xspace}
\newcommand{\emccd}{electron multiplying CCD (EMCCD)\renewcommand{\emccd}{EMCCD\xspace}\xspace}
\newcommand{\dasp}{the Durham \ao simulation platform (DASP)\renewcommand{\dasp}{DASP\xspace}\xspace}
\newcommand{\eelt}{European \elt (E-ELT)\renewcommand{\eelt}{E-ELT\xspace}\xspace}
\newcommand{\mpi}{Message Passing Interface (MPI)\renewcommand{\mpi}{MPI\xspace}\xspace}
\newcommand{\smp}{symmetric multi-processing (SMP)\renewcommand{\smp}{SMP\xspace}\xspace}
\newcommand{\svd}{singular value decomposition (SVD)\renewcommand{\svd}{SVD\xspace}\xspace}
\newcommand{\gpu}{graphical processing unit (GPU)\renewcommand{\gpu}{GPU\xspace}\renewcommand{\gpus}{GPUs\xspace}\xspace}
\newcommand{\gpus}{graphical processing units (GPUs)\renewcommand{\gpu}{GPU\xspace}\renewcommand{\gpus}{GPUs\xspace}\xspace}
\newcommand{\fft}{fast Fourier transform (FFT)\renewcommand{\fft}{FFT\xspace}\xspace}
\newcommand{\ifu}{integral field unit (IFU)\renewcommand{\ifu}{IFU\xspace}\xspace}
\newcommand{\darc}{the Durham adaptive optics real-time controller (DARC)\renewcommand{\darc}{DARC\xspace}\renewcommand{\Darc}{DARC\xspace}\xspace}
\newcommand{\Darc}{The Durham adaptive optics real-time controller (DARC)\renewcommand{\darc}{DARC\xspace}\renewcommand{\Darc}{DARC\xspace}\xspace}
\newcommand{\cots}{commercial off-the-shelf (COTS)\renewcommand{\cots}{COTS\xspace}\xspace}
\newcommand{\rtcp}{real-time control pipeline (RTCP)\renewcommand{\rtcp}{RTCP\xspace}\xspace}
\newcommand{\rms}{root-mean-square (RMS)\renewcommand{\rms}{RMS\xspace}\xspace}
\newcommand{\sFPDP}{serial Front Panel Data Port (sFPDP)\renewcommand{\sFPDP}{sFPDP\xspace}\xspace}
\newcommand{\wpu}{wavefront processing unit (WPU)\renewcommand{\wpu}{WPU\xspace}\xspace}

\newcommand{\rtcs}{real-time control system (RTCS)\renewcommand{\rtcs}{RTCS\xspace}\xspace}
\newcommand{\ptp}{point-to-point (PTP)\renewcommand{\ptp}{PTP\xspace}\xspace}
\newcommand{\sse}{streaming SIMD extension (SSE)\renewcommand{\sse}{SSE\xspace}\xspace}
\newcommand{\api}{application programming interface (API)\renewcommand{\api}{API\xspace}\xspace}
\newcommand{\corba}{Common Object Request Broker Architecture (CORBA)\renewcommand{\corba}{CORBA\xspace}\xspace}
\newcommand{\lqg}{linear quadratic gaussian (LQG)\renewcommand{\lqg}{LQG\xspace}\xspace}
\newcommand{\scao}{single conjugate adaptive optics (SCAO)\renewcommand{\scao}{SCAO\xspace}\xspace}
\newcommand{\dma}{direct memory access (DMA)\renewcommand{\dma}{DMA\xspace}\xspace}
\newcommand{\xao}{extreme adaptive optics (XAO)\renewcommand{\xao}{XAO\xspace}\xspace}
\newcommand{\vlt}{Very Large Telescope (VLT)\renewcommand{\vlt}{VLT\xspace}\xspace}
\newcommand{\sparta}{Standard Platform for Advanced Real-Time
  Applications (SPARTA)\renewcommand{\sparta}{SPARTA\xspace}\xspace}
\newcommand{\eso}{European Southern Observatory (ESO)\renewcommand{\eso}{ESO\xspace}\xspace}
\newcommand{\eagle}{EAGLE\xspace}
\newcommand{\epics}{Exo-Planet Imaging Camera and Spectrograph (EPICS)\renewcommand{\epics}{EPICS\xspace}\xspace}
\newcommand{\iir}{infinite impulse response (IIR)\renewcommand{\iir}{IIR\xspace}\xspace}
\newcommand{\gtc}{Gran Telescopio Canarias (GTC)\renewcommand{\gtc}{GTC\xspace}\xspace}
\newcommand{\cog}{centre of gravity (CoG)\renewcommand{\cog}{CoG\xspace}\xspace}

\title[ELT MOAO DM fault tolerance]{Faulty actuator tolerance in deformable
  mirrors for Extremely Large Telescope multi-object adaptive optics}
\author[A. G. Basden]{A. G. Basden$^{1}$\thanks{E-mail:
    a.g.basden@durham.ac.uk (AGB)}\\
$^{1}$Department of Physics, South Road, Durham, DH1 3LE, UK}

\begin{document}
\maketitle

\begin{abstract}
Planned instruments utilising multi-object adaptive optics systems on
the forthcoming extremely large telescopes require large numbers of
high order deformable mirrors.  These devices are a significant cost
driver, particularly if specifications regarding the number of faulty
actuators are stringent.  Here, we investigate the effect on adaptive
optics performance that such faulty actuators have, and draw
conclusions about how far faulty actuator specifications (and hence
cost) can be relaxed without having a significant effect on adaptive
optics performance.  We also provide performance estimates using a map
of faulty actuators from an existing deformable mirror.  We
investigate the effect of faulty actuators using an end-to-end Monte
Carlo adaptive optics simulation code.  We find that for actuators
stuck at a fixed height above the deformable mirror surface, between
1--2\% of actuators can be faulty before significant performance
degradation occurs.  For actuators that a coupled to nearest
neighbours, up to about 5\%, can be faulty before \ao performance
begins to be affected.
\end{abstract}
\begin{keywords}
Instrumentation: adaptive optics,
instrumentation: high angular resolution,
Methods: numerical
\end{keywords}

\section{Introduction}
The forthcoming generation of optical ground-based \elts will have
primary mirror diameters of over 30~m.  These facilities will depend
on advanced \ao \citep{adaptiveoptics} to meet the majority of their
scientific goals, and will provide astronomers with the necessary
resolutions and light collecting areas to probe the universe with
unprecedented sensitivity.  Instruments using advanced modes of \ao
operation are proposed, including \moao and \mcao which will deliver
\ao corrected \psfs over a wide field-of-view.  These systems require
more than one \dm, and \moao systems in particular require a large
number of \dms, one for each corrected science channel.  Designs for
the \eelt \eagle instrument \citep{2010SPIE.7736E..25Rshort} include
20 separately corrected \ao channels, thus requiring 20 \dms.

The cost of \dms are significant factor for consideration in the
design of \moao systems, and any reduction in cost would be welcomed
by the design teams.  The failure rate of individual \dm actuators
can also be a risk to these instruments.  Here we investigate the
effect of the presence of faulty \dm actuators on \moao performance
using a full end-to-end Monte-Carlo simulation code, \dasp
\citep{basden5}.  We model an instrument similar to \eagle, and
consider actuators stuck at different surface heights, including at
the full range of the \dms, and also consider actuators that are
coupled to nearest neighbours.

In \S2, we describe our simulation models and assumptions.  In \S3, we
present our results and findings, and we conclude in \S4.

\section{Simulation of deformable mirror faults}
The phase A design of the \eagle instrument includes six \lgss and up to five
\ngss, which will be used to provide tomographic wavefront information
over a ten arc-minute field of view.  Twenty science channel pick-off
mirrors will be used to direct small regions of interest into integral
field spectrographs via individual \dms, which provide wavefront
correction for that part of the field of view.  The cost and
availability of these \dms represents a major risk for \eagle, driven
by the large number required, and the high order of the \eelt \wfss.
A previous study \citep{basden12} has shown that when the \eelt M4 \dm
(the fourth mirror in the telescope optical train) is taken into
account, the individual \eagle \dms can have their order (number of
actuators) reduced to $64\times64$ without significantly reducing \ao
performance, thus bringing them into the regime of \dms that are
currently commercially obtainable.

Unfortunately, it has not yet been possible to manufacture suitable
\dms of this order that are free of defects.  Indeed, a stringent
requirement for a low number of defects significantly adds to the \dm
cost and may require a development effort, since currently, a large
number of samples must be manufactured and the best selected
\citep{doi:10.1117/12.811681,doi:10.1117/12.687687}.  By reducing \dm
specification by allowing a greater number of faulty actuators, it
will be possible to reduce the overall cost of the \eagle \dms or the
risk associated with them, and here we provide details of such a
study.

\subsection{Simulation model details}
We model \dms with faulty actuators in a full end-to-end \ao
simulation, including atmosphere and telescope
models, to allow the effect of faulty actuators on expected \eagle \ao
performance to be investigated.  We perform full Monte-Carlo
simulation of the atmosphere, the telescope, wavefront sensors and
deformable mirrors in the results presented here.

We model the atmosphere using a standard \eso nine
layer turbulence profile \citep{miskaltao} as given in
table~\ref{tab:atmos}, an outer scale of 25~m, and a Fried's parameter
of 13.5~cm.

\begin{table*}
\centering
\begin{minipage}{150mm}
\begin{tabularx}{\linewidth}{Xccccccccc}
C$_n^2$ profile & Layer 1 & Layer 2 & Layer 3 & Layer 4 & Layer 5 &
  Layer 6 & Layer 7 & Layer 8 & Layer 9 \\ \hline
Height / m & 47 & 140 & 281 & 562 & 1125 & 2250 & 4500 & 9000 & 18000\\
C$_n^2$ / \% & 52.24 & 2.6 & 4.44 & 11.6 & 9.89 & 2.95 &
5.98 & 4.3 & 6 \\
Speed / ms$^{-1}$ & 4.55 & 12.61 & 12.61 &  8.73 &  8.73 & 14.55 &
24.25 & 38.8 & 20.37\\
Direction / $^\circ$ & 0 &  36 &  72 & 108 & 144 & 180 & 216 & 252 & 288\\
\end{tabularx}
\caption{A table giving the atmospheric layer heights above the
  primary mirror, and corresponding layer strengths used in the
  simulations here, taken from \citet{miskaltao}.}
\label{tab:atmos}
\end{minipage}
\end{table*}

We assume a telescope with an outer diameter of 39.3~m, and a 11.2~m
central obscuration.  For the results presented here we use \lgs
tomography only and assume that the \lgs tip-tilt measurements are
valid.  This removes any uncertainly on \ngs availability and asterism
details, which for the purposes of this study are irrelevant.  We
assume $80\times80$ sub-apertures for each wavefront sensor, with a
matching pitch for the M4 \dm which is global to all science channels.
The individual \moao \dms have $64\times64$ actuators unless otherwise
stated.  The \lgss are arranged regularly on a circle with a
220~arc-second radius.  Our \lgs spots are elongated to model a sodium
layer at 90~km with a 10~km full-width at half-maximum, and we include
the cone effect in our simulations.  We use $10^6$ photons per
sub-aperture per \wfs frame so that we are working in the high-light
level regime, and photon shot noise is included.  We measure science
performance on-axis at 1.65~$\mu$m, and have previously shown
\citep{basden12} that performance is relatively uniform across the
5~arc-minute science field.  Unless otherwise stated, performance is
given as the percentage of ensquared energy within 75~mas, which for
\eagle is a key performance criteria.

Tomographic wavefront reconstruction is performed at each of the nine
atmospheric layers, with a Laplacian regularisation to approximate
wavefront phase covariance.  We assume a \wfs frame rate of 250~Hz and
ensure that the science \psfs are well averaged.

\subsection{Deformable mirror fault modelling}
We model our \dms using a cubic spline interpolation function, which
takes actuator positions and heights as an input, and provides an
interpolated, higher spatial resolution output that is a model of the
mirror surface.  In these investigations of \dm faults, we have
several cases to investigate:

\begin{enumerate}
\item Particular actuators are stuck at some value, with an otherwise
  ideal \dm.
\item Particular actuators are stuck at some value, with a \dm that
  has a limited total stroke.
\item Particular actuators are stuck at the \dm stroke limits.
\item Stuck actuators can be spread individually over the \dm surface,
  or clumped together in groups.
\item Actuators can be coupled to nearest neighbours, with a motion
  equal to the average of nearest neighbours.
\item A fraction of actuators stuck below the mean \dm surface, and
  another fraction stuck above the mean \dm surface.
\end{enumerate}

We also consider cases where the global M4 \dm is not present, as a
risk mitigation, to provide performance estimates should this \dm not
be operational, and for completeness so that results can be translated
to other systems more easily.

\section{Effect of DM faults on AO performance}
We generate an actuator map with faulty actuators in random positions
across the \dm surface (which, for a given random number seed, is
repeatable).  We only count faulty actuators that are within the
telescope pupil function, i.e.\ those obscured by the central
obscuration, or at the \dm corners are not counted.  We then
investigate \ao performance as a function of the number of such faulty
actuators.

\subsection{Actuators stuck at mid-range}
\begin{figure}
\includegraphics[width=\linewidth]{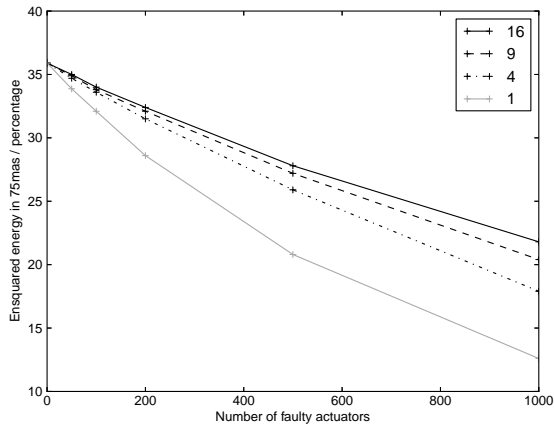}
\caption{A figure showing \ao performance as a function of number of
  actuators stuck at mid-range position.  Individual faulty actuators,
  and groupings of faulty actuators ($2\times2$, $3\times3$ and
  $4\times4$) are shown.}
\label{fig:midrange}
\end{figure}

Depending on \dm technology, the most likely positions for actuators
to stick at may be at the mid-range position, at full stroke (in
either direction), or at some other position.  Fig.~\ref{fig:midrange}
shows \ao performance as a function of number of actuators stuck at
the \dm mid-range position, i.e. the position that defines a
zero-point for the \dm surface.  Uncertainties are not shown, but are
approximately 0.05\%, obtained by multiple realisations with different
random number seeds.

We consider the case where individual actuators are faulty, and where
clumps of actuators are faulty, and investigate how performance is
affected by the size of these clumps.  From this figure, we can see
that having small clumps of faulty actuators is less harmful to \ao
performance than having the same number of individual faulty actuators
(i.e.\ the total number of faulty actuators is the same in each case).

For the remainder of this paper we consider only the case for
individual actuators, i.e.\ we do not consider faulty actuators
grouped together in clumps.

\subsection{Actuators stuck at other surface heights}
\begin{figure}
\includegraphics[width=\linewidth]{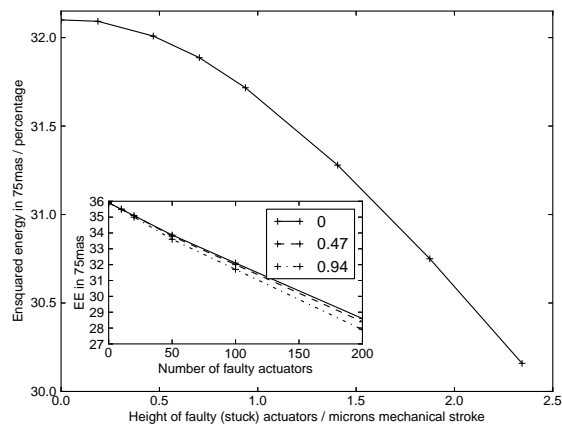}
\caption{A figure showing \ao performance as a function of the height
  at which 100 faulty actuators are stuck, above the \dm mean surface
  position.  Inset is shown for the \ao performance as a function of the number
  faulty of actuators stuck at a given height above the mean \dm
  surface, with the legend giving the height above mean surface position
  at which actuators are stuck in microns (mechanical).  Total \dm
  stroke is unlimited here.}  
\label{fig:stuck}
\end{figure}

We consider the case where \dm actuators are stuck at some height
other than the mean phase position, for \dms both with and without
limited stroke.  Fig.~\ref{fig:stuck} shows that there is only a
slight performance dependence on the height at which actuators are
stuck (for an otherwise unlimited \dm) when this height is small.
However, as the faulty actuator height increases, performance is
degraded since these actuators have a larger impact on corrected
wavefront phase.  For \dms with actuators stuck at more than one
surface height, we find that performance is almost identical to the
case where actuators are stuck at a single height (here we consider
only the case with actuators stuck at equal positive or negative
surface heights).

\begin{figure}
\includegraphics[width=\linewidth]{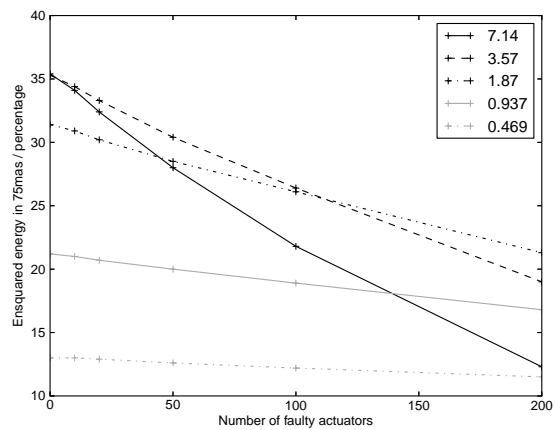}
\caption{A figure showing \ao performance as a function of number of
  actuators stuck at the stroke limits of the \dm.  The stroke limit
  of the \dm is given in the key in microns of mechanical stroke.}
\label{fig:stroke}
\end{figure}

We have previously shown \citep{basden12} that limited \moao \dm
stroke affects \ao performance, and that, as expected, \dms with larger
stroke provide better performance.  However, in the case of a \dm with
faulty actuators which are stuck at the limits of the \dm surface, it
is evident that large stroke will no longer give best performance,
since this would result in actuators stuck at high surface heights.
Rather, as shown in Fig.~\ref{fig:stroke}, there is a trade-off
between the effect of stuck actuators, and the effect of actuator
clipping on working actuators when they reach the \dm stroke range.
It should be noted that the optimal stroke will change under different
atmospheric conditions.  Worse seeing will require a larger \dm stroke
to avoid performance degradation, while in better seeing, if the
stroke is too large then the faulty actuators will have a larger
effect on performance.  

It is interesting to note that if there are 200 faulty actuators stuck
at the stroke limit, then a \dm with 7~$\mu$m stroke will offer worse
performance than one with about 1$~\mu$m stroke.  Therefore, care
should be taken when specifying maximum stroke and faulty actuator
requirements.  When actuators are stuck at the \dm stroke limit, it
should be noted that performance drops much more quickly with number
of faulty actuators than when they are stuck at mid-range.  We suggest
that when actuators are stuck at \dm stroke limits, only 10-20
actuators (about 0.2\%) should be accepted before \ao corrected
ensquared energy falls by more than 1\%.

\begin{figure}
\includegraphics[width=\linewidth]{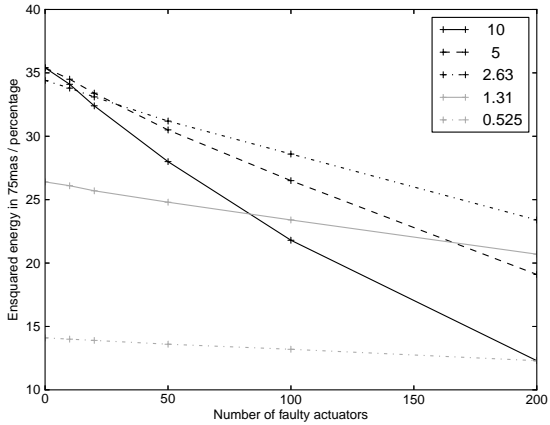}
\caption{A figure showing \ao performance as a function of number of
  actuators stuck at a height of about 70\% of the stroke limits of
  the \dm.  The stroke limit of the \dm is given in the key in microns
  of mechanical stroke.}
\label{fig:fracstroke}
\end{figure}

Fig.~\ref{fig:fracstroke} shows \ao performance as a function of
number of faulty actuators, which are stuck at about 70\% of the
maximum \dm stroke, which is given in the legend.  This again shows
that lower total \dm stroke can be an advantage when faulty actuators
are present, and covers the case where faulty actuators are stuck at
some level not equal to the full \dm stroke limits.
\ignore{actually at 71.393939 = 589/1650.*2}

\subsection{Actuator coupling}
\begin{figure}
\includegraphics[width=\linewidth]{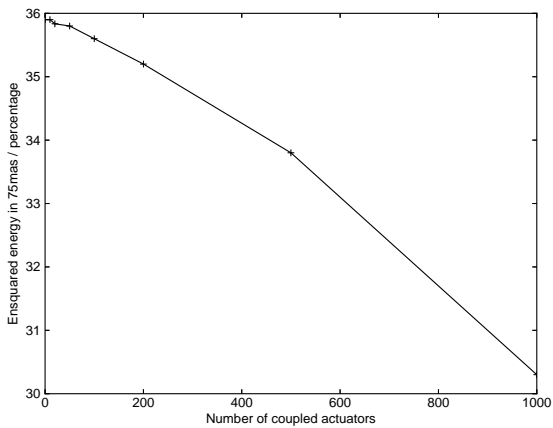}
\caption{A figure showing \ao performance as a function of number of
  coupled actuators.}
\label{fig:coupled}
\end{figure}

We have investigated cases where a fraction of actuators are not
directly controlled, but move to the mean position of their
neighbours.  Fig.~\ref{fig:coupled} shows \ao performance as a
function of number of coupled actuators.  For \eagle, there is little
degradation in performance until over 100 actuators are coupled in
this way, and that even when 5\% of actuators are coupled, ensquared
energy within 75~mas falls by less than 1\%.  As would be expected,
having a given number of coupled actuators is far less harmful to \ao
performance than having the same number of stuck actuators.

\subsection{Performance without the global M4 DM}
\begin{figure}
\includegraphics[width=\linewidth]{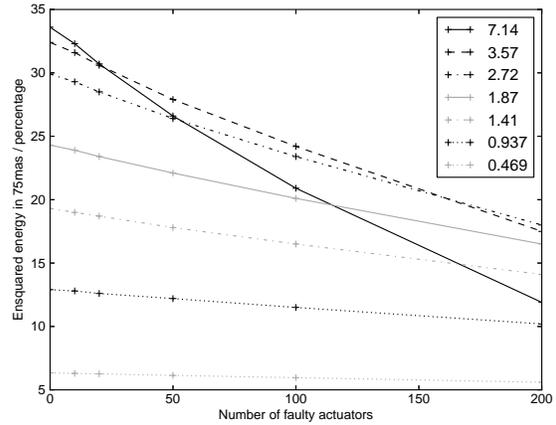}
\caption{A figure showing \ao performance as a function of number of
  actuators stuck at the stroke limits of the \dm.  In this case,
  there is no M4 mirror present, and all correction is performed by
  the \moao \dm.  The mechanical stroke limits of the \dm are given in the
  legend, in microns.}
\label{fig:nom4}
\end{figure}

We now consider (partly for completeness, partly for risk mitigation)
the case where the \eelt M4 \dm is not used for \ao correction.
Therefore, the \moao \dms must provide all of the correction.
Fig.~\ref{fig:nom4} shows performance as a function of number of
actuators stuck at the \dm stroke limit, for \dms with different total
stroke capability.

\begin{figure}
\includegraphics[width=\linewidth]{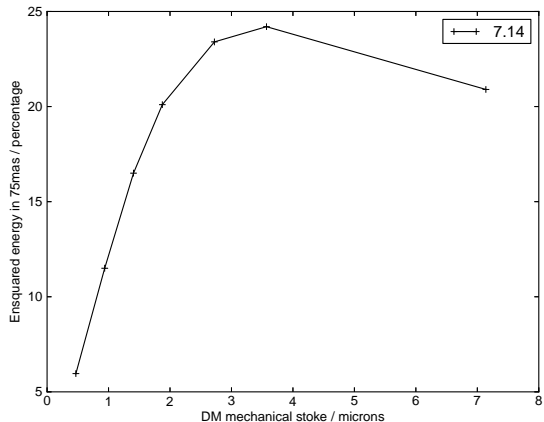}
\caption{A figure showing \ao performance as a maximum \dm stroke when
  100 actuators are stuck at the limit of the \dm stroke.  In this
  case, there is no M4 mirror present, and all correction is performed
  by the \moao \dm.}
\label{fig:nom4at100}
\end{figure}

It can be seen that when there are no faulty actuators, best
performance is obtained when \dm stroke is large, as expected.
However, when stroke is large, \ao performance falls faster once the
number of faulty actuators increases.  This is an important
consideration for \ao system designers: If it is expected that there
will be a number of faulty actuators, stuck at the stroke limits of
the \dm, then it may be better to choose a \dm with reduced stroke.
As can be seen from Fig.~\ref{fig:nom4at100}, if there are 100 faulty
actuators, a \dm with stroke limited to 2.7~$\mu$m or 3.5~$\mu$m gives better
performance than one limited to about 7~$\mu$m.

The explanation for this is simple:  An increased number of faulty
actuators will result in increased phase distortion if these actuators
are stuck further from mean position, worsening performance.

In the case where all actuators are stuck in one direction (i.e.\ all
high, or all low), rather than having some actuators stuck high, and
some stuck low, some risk mitigation can be made with a large stroke
\dm, by altering the defined mid-point, i.e.\ by moving the \dm
zero-point closer to the stroke limits at which the actuators are
stuck.  This effectively clips the \dm more stringently in one
direction.  In this case, a \dm with higher stroke will always yield
better performance.  However, this can come at the expense of some
complexity, since \dms are not always as well behaved further from
their mid-point, and can have linearity and surface flatness issues.

\subsection{Modelling of a known faulty actuator map}
To add a degree of realism to our simulations, we consider the case of
the dead actuator map for an existing $64\times64$ actuator \dm, that
is used by the Gemini Planet Imager instrument
\citep{2012SPIE.8446E..1UMshort}.  This map includes actuators that
are stuck in both directions, and also actuators that are coupled to
nearest neighbours.  A total of 23 actuators were coupled (about 0.5\%
of all available actuators), five stuck high and three stuck low (a
total of about 0.2\% of available actuators).  At this level, we find
that \ao performance is almost no different from when using a perfect
\dm, as can be expected by considering cases in previous sections with
similar numbers of faulty actuators.

It is likely that for an \elt instrument such as \eagle, a larger
number of faulty actuators than this will be present, since the
requirement for a large number of \dms at low cost means that
imperfections may have to be accepted unless there are improvements in
\dm manufacturing techniques.  It is not unknown for \dm actuators to
fail with age, which will also increase the numbers of faulty
actuators.  However, as the results in previous sections show, the
performance of \elt instruments such as \eagle only becomes
significantly affected when 1--2\% of actuators, or more in the case
of coupled actuators (up to 5\%), are faulty.  The \eagle instrument
can therefore cope with a larger
fraction of faulty actuators than evident on an existing \dm, allowing
some of the risk associated with \eagle to be removed.

We have discussed cost reductions due to accepting increased numbers
of faulty actuators with a leading \dm manufacturer, and such
reductions would be possible, though we do not discuss this further here.

\section{Conclusions}
We have made a study of the effect of faulty \dm actuators on
\elt-scale \moao performance using an end-to-end Monte-Carlo
simulation tool.  Individual \dm actuators that are stuck at some
level relative to the mean surface position have been considered, as
well as actuators which are coupled to nearest neighbours.  For the
instrument description considered here, we find that between 1--2\% of
actuators can be faulty (stuck at mid-range) before significant
performance degradation (of more than a 1--2\% drop in ensquared
energy) occurs.  When actuators are stuck at the stroke limit of
the \dm, the requirements are more stringent, with only 0.2\% of
actuators allowed to be faulty before ensquared energy drops by more
than 1\%.  We find that care should be taken when specifying maximum \dm
stroke if it is known that faulty (stuck) actuators will be present.

For actuators that a coupled to nearest neighbours, a greater
fraction, up to about 5\%, can be faulty before \ao performance begins
to be affected.  It should be noted that more stringent requirements
will be needed for instruments demanding higher \ao performance, such
as \ltao or \xao systems. 

These findings allow the risk and cost for an instrument such as
\eagle to be significantly reduced, and can be applied to any
instrument requiring a large number of high order \dms that are
currently at the cutting edge of technological innovation.

\section*{Acknowledgements}
This work is funded by the UK Science and Technology Facilities
Council, grant ST/K003569/1.

\bibliographystyle{mn2e}

\bibliography{mybib}
\bsp

\end{document}